\documentclass{PoS}

\usepackage{url}
\usepackage{color}

\title{Neutron-skin effect and centrality dependence of high-$p_{\mathrm{T}}$ observables in nuclear collisions}

\ShortTitle{Neutron-skin effect and centrality dependence of $p_{\mathrm{T}}$ observables in nuclear collisions}

\author{\speaker{Ilkka Helenius}$^{a}$, Hannu Paukkunen$^{bcd}$ and Kari~J. Eskola$^{bc}$\\
\llap{$^a$}Department of Astronomy and Theoretical Physics, Lund University, S\"{o}lvegatan 14A, SE-223 62 Lund, Sweden\\
\llap{$^b$}University of Jyvaskyla, Department of Physics, P.O. Box 35, FI-40014 University of Jyvaskyla, Finland\\
\llap{$^c$}Helsinki Institute of Physics, P.O. Box 64, FI-00014 University of Helsinki, Finland\\
\llap{$^d$}Departamento de F\'{\i}sica de Part\'{\i}culas and IGFAE, Universidade de Santiago de Compostela, E-15782 Galicia, Spain\\
E-mail: \email{ilkka.helenius@thep.lu.se}, \email{hannu.t.paukkunen@jyu.fi}, \email{kari.eskola@jyu.fi}}

\abstract{
We report on our studies of the neutron-skin effects in high-$p_{\rm T}$ observables at the LHC. We study the impact of the neutron-skin effect on the centrality dependence of inclusive direct photon, high-$p_{\rm T}$ hadron and $W^{\pm}$ production in nuclear collisions at the LHC. The neutron-skin effect refers to the observation that in spherical heavy nuclei, the tail of the neutron distribution extends farther than the distribution of protons, which can affect observables sensitive to electroweak phenomena in very peripheral collisions. We quantify this effect for direct photons, charged hadrons and W bosons as a function of the collision centrality. In the case of direct photons we find that it will be difficult to resolve the neutron-skin effect, given the uncertainties in the nuclear PDFs and their spatial dependence. With charged hadrons and W's, however, up to 20~\% unambiguous effects are expected for most peripheral collisions.}

\FullConference{XXIV International Workshop on Deep-Inelastic Scattering and Related Subjects\\
         11-15 April, 2016\\
         DESY Hamburg, Germany}

\begin{document}

\section{Introduction}

The geometry of a nuclear collision is determined by the impact parameter, the distance between the centers of the colliding nuclei. The impact parameter cannot be directly measured experimentally but a correlation between the event activity and the impact parameter is expected. Typically the energy measured in a given part of a detector is used to classify the events in terms of centrality. Theoretically, the corresponding centrality classification can be done either with the Optical Glauber model, where the centrality classes are defined as impact parameter intervals, or with the Monte-Carlo Glauber, which takes into account the fluctuations in the nucleon configurations and uses the number of binary collisions as a measure of centrality. In principle it is by no means guaranteed that the experimental and theoretical classifications would be equivalent, but in the case of heavy-ion collisions (e.g. Pb+Pb) the theoretical and experimental centrality classifications are found to produce similar results for many observables.  In a collision of a small projectile and a large target (e.g. p+Pb), however, the experimental centrality measures have been noticed to induce a non-trivial correlation between the centrality determination and the high-$p_{\rm T}$ observable. 

In this talk we quantify the impact of the neutron-skin (NS) effect on the centrality dependence of inclusive direct photon, charged hadron and $W^{\pm}$ production in nuclear collisions at the LHC. The NS effect refers to the observation at MeV-scale experiments that the tail of the neutron density distribution extends farther than in the case of the proton density \cite{Tarbert:2013jze}. This influences the centrality dependence of the observables sensitive to electroweak couplings. The results presented here are from Refs.~\cite{Paukkunen:2015bwa, HPE2016}. Once confirmed in high-energy experiments, the NS effect could help, in turn, to also calibrate the centrality classification methods at the LHC.

\section{Framework}

Both the Optical and Monte-Carlo Glauber models require the nuclear density distribution as an external input. The density can be described with the two-parameter Fermi (2pF) distribution
\begin{equation}
\rho^{A}(\mathbf{r}) = \rho_0^{A}/(1 + \mathrm{e}^{((|\mathbf{r}| - R_{A})/d_{A})}),
\label{eq:2pF}
\end{equation}
where the parameter $R_{A}$ describes the nuclear core size, $d_{A}$ the thickness of the nuclear skin, and $\rho_0^{A}$ is fixed by the normalization $\int \mathrm{d}^3\mathbf{r}\,\rho^{A}(\mathbf{r})=A$, where $A$ is the nuclear mass number.  Canonically the parameters in 2pF distribution are taken to be the same for protons and neutrons. However, when the NS effect is taken into account, different parameter values for proton and neutron density should be used. We use the values from Ref.~\cite{Tarbert:2013jze} which are listed in Table \ref{tab:parameters}. 
\begin{table}[htb]
\begin{center}
\begin{tabular}{lcc}
 & proton & neutron \\
\hline
 $R_{\mathrm{Pb}}$ & $6.680~\mathrm{fm}$ & $6.70\pm 0.03~\mathrm{fm}$ \\ 
 $d_{\mathrm{Pb}}$ & $0.447~\mathrm{fm}$ & $0.55\pm 0.03~\mathrm{fm}$
\end{tabular}
\end{center}
\caption{The parameter values for the 2pF distribution for protons and neutrons according to Ref.~\cite{Tarbert:2013jze}.}
\label{tab:parameters}
\end{table}

Here, the centrality classification is done using Optical Glauber model where the total inelastic cross section in a collision of nuclei $A$ and $B$ can be calculated from
\begin{equation}
\sigma_{AB}^{\mathrm{inel}}(\sqrt{s_{\mathrm{NN}}}) = \int \mathrm{d}^2\mathbf{b} \left[ 1 - \mathrm{e}^{-T_{AB}(\mathrm{b})\,\sigma_{\mathrm{NN}}^{\mathrm{inel}}(\sqrt{s_{\mathrm{NN}}})} \right],
\end{equation}
where the nuclear overlap function is obtained from 
\begin{equation}
T_{AB}(\mathrm{b}) = \int \mathrm{d}^2\mathbf{s} \left[ T_A^{\mathrm{p}}(\mathbf{s_1}) + T_A^{\mathrm{n}}(\mathbf{s_1})\right] \left[ T_B^{\mathrm{p}}(\mathbf{s_2}) + T_B^{\mathrm{n}}(\mathbf{s_2})\right].
\end{equation}
The transverse vectors are defined as $\mathbf{s_1} = \mathbf{s}+\mathbf{b}/2$ and $\mathbf{s_2} = \mathbf{s}-\mathbf{b}/2$, and the nuclear thickness for nucleons of type $i$ is obtained from $T_A^i(\mathbf{s}) = \int \mathrm{d}z \, \rho^{i,A}(z, \mathbf{s})$. The $\sigma_{\mathrm{NN}}^{\mathrm{inel}}(\sqrt{s_{\mathrm{NN}}})$ gives the total inelastic nucleon-nucleon cross section at a given collision energy $\sqrt{s_{\mathrm{NN}}}$ for which we use values $\sigma_{\mathrm{NN}}^{\mathrm{inel}}(\sqrt{s_{\mathrm{NN}}}=2.76~\mathrm{TeV}) = 65~\mathrm{mb}$ and $\sigma_{\mathrm{NN}}^{\mathrm{inel}}(\sqrt{s_{\mathrm{NN}}}=5.0~\mathrm{TeV}) = 70~\mathrm{mb}$ from Ref.~\cite{Antchev:2013iaa}. The impact parameters corresponding to a given centrality class $\mathcal{C}_k$ are then obtained by requiring that the integral over the range $[b_k,b_{k+1}]$ yields a certain fraction of the total inelastic cross section.

The hard-process cross section in a given centrality class is then computed by
\begin{equation}
\mathrm{d}\sigma_{AB}^{\mathrm{hard}}(\mathcal{C}_k) = 2\pi \int_{b_{k}}^{b_{k+1}} \mathrm{d}b\, b \int \mathrm{d}^2\mathbf{s} \sum_{i,j} T_A^i(\mathbf{s_1})T_B^j(\mathbf{s_2}) \mathrm{d}\sigma_{ij}^{\mathrm{hard}}(A,B,\mathbf{s_1}, \mathbf{s_2}),
\end{equation}
where the indices $i$ and $j$ run over different combinations of protons and neutrons. The hard nucleon-nucleon cross section $\sigma_{ij}^{\mathrm{hard}}(A,B,\mathbf{s_1}, \mathbf{s_2})$ is calculated as a convolution between the nuclear parton distribution functions (nPDFs) and the partonic coefficient functions. The spatial dependence of $\mathrm{d}\sigma_{ij}^{\mathrm{hard}}(A,B,\mathbf{s_1}, \mathbf{s_2})$ follows from the spatial dependence of the nPDFs, which we take from \textsc{eps09s} \cite{Helenius:2012wd, Eskola:2009uj}. The needed free nucleon PDFs are from the \textsc{ct10} analysis \cite{Lai:2010vv}.


\section{Direct photons}

\begin{figure}[thb]
\centering
\includegraphics[width=0.49\textwidth]{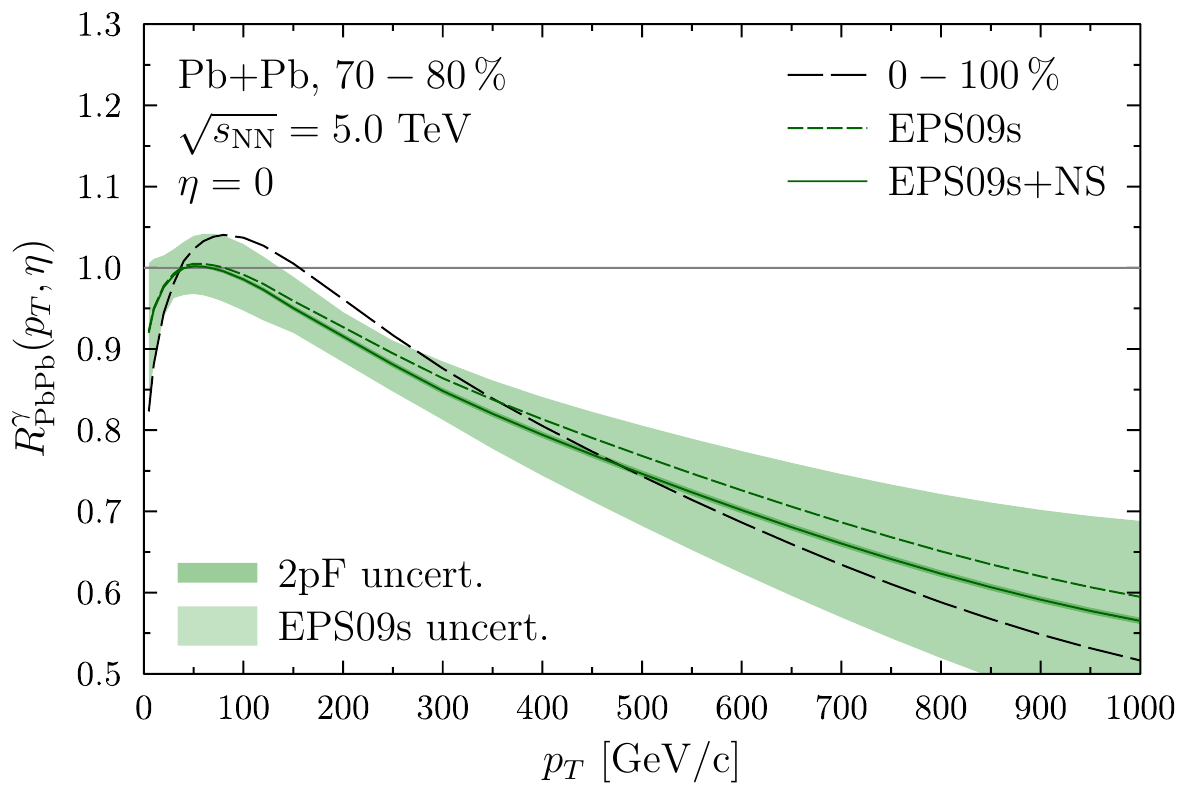}
\includegraphics[width=0.49\textwidth]{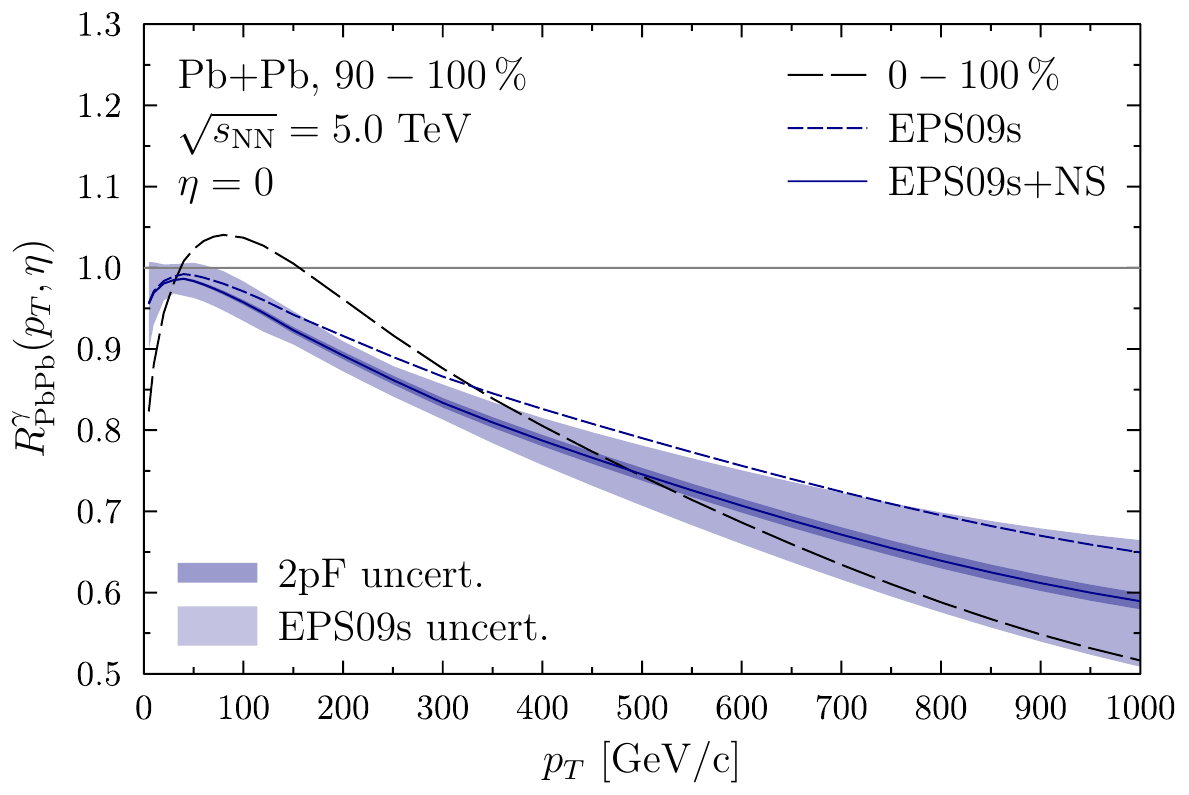}
\caption{The nuclear modification factor for inclusive direct photons in Pb+Pb collisions at $\sqrt{s_{\mathrm{NN}}} = 5.0~\mathrm{TeV}$ and $\eta = 0$ for 70--80~\% (left) and 90--100~\% (right) centrality classes. The results with the NS effect (solid) are compared to results without the NS effect (dashed) and to MB results (long-dashed). The uncertainty bands originate from nPDF uncertainties (light colour band) and from the 2pF distribution (dark colour band). Based on Ref.~\cite{HPE2016}.}
\label{fig:RPbPb}
\end{figure}

The cross section for inclusive direct photon production is calculated at next-to-leading order (NLO) using the \textsc{Incnlo} code \cite{incnlopage} and applying the BFG-II parton-to-photon fragmentation functions (FFs) \cite{Bourhis:1997yu}. Figure~\ref{fig:RPbPb} shows the nuclear modification factor for the direct photons at 70--80~\% and 90--100~\% centralities with and without the NS effect. Also the minimum bias (MB) result (0--100~\%) is plotted for comparison. The uncertainty bands are derived from the nPDF error sets and from the uncertainties in the 2pF distribution. The NS effect has a larger impact at high-$p_{\rm T}$ where the sensitivity to the valence quarks gets increased. At the highest $p_{\rm T}$ values the NS effect generates a few-percent suppression for the 70--80~\% centrality due to the neutron-involving collisions in which the valence quark carry less charge (in comparison to the valence quarks of the proton), and a bit stronger suppression for the 90--100~\% class. The uncertainty arising from the 2pF distribution is very small in the 70--80~\% bin since the ``valence-quark-from-proton'' contribution dominates the cross section which is not directly sensitive to the neutron 2pF-distribution variations. In the 90--100~\% bin the 2pF uncertainty is larger since the relative amount of protons becomes smaller at high impact parameters. However, the nPDF uncertainties are still larger than the NS effect, and since the centrality dependence of the nPDFs has an opposite effect than the NS effect, it is difficult to distinguish the NS effect with direct photons. Some of the nPDF uncertainties cancel out when considering central-to-peripheral ratio, see Ref.~\cite{HPE2016}.

\section{Charged hadrons}

In the case of charged hadrons we consider the ratio between the cross sections for negative and positive high-$p_{\rm T}$ hadrons, $h^-/h^+$. The cross sections are calculated by convoluting the NLO partonic spectra with parton-to-hadron FFs using the \textsc{incnlo} code. Two FF analyses are considered, \textsc{dss} \cite{deFlorian:2007hc} and \textsc{kretzer} \cite{Kretzer:2000yf}. Here, we do not try to model possible final state effects as we are mainly interest in the region $p_{\rm T} \gg 100~\mathrm{GeV/c}$ where e.g. energy-loss phenomena can be expected to become negligible \cite{CMS:2016ouo}. In addition, at lower $p_{\rm T}$, where the final state effect are important, the kaon-to-pion and proton-to-pion ratios have been observed to be the same in Pb+Pb and p+p collisions \cite{Adam:2015kca}, suggesting that the final state interactions are not important in the ratio $h^-/h^+$.

\begin{figure}[thb]
\centering
\includegraphics[width=0.49\textwidth]{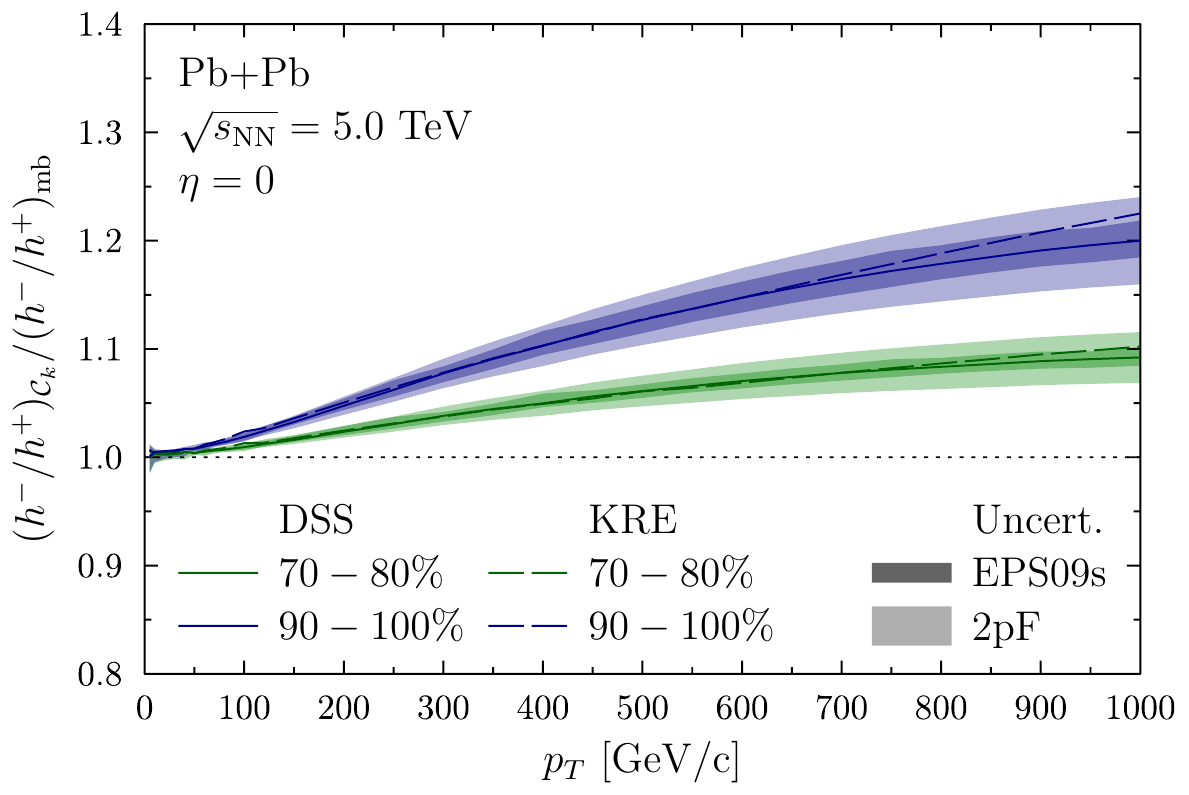}
\includegraphics[width=0.49\textwidth]{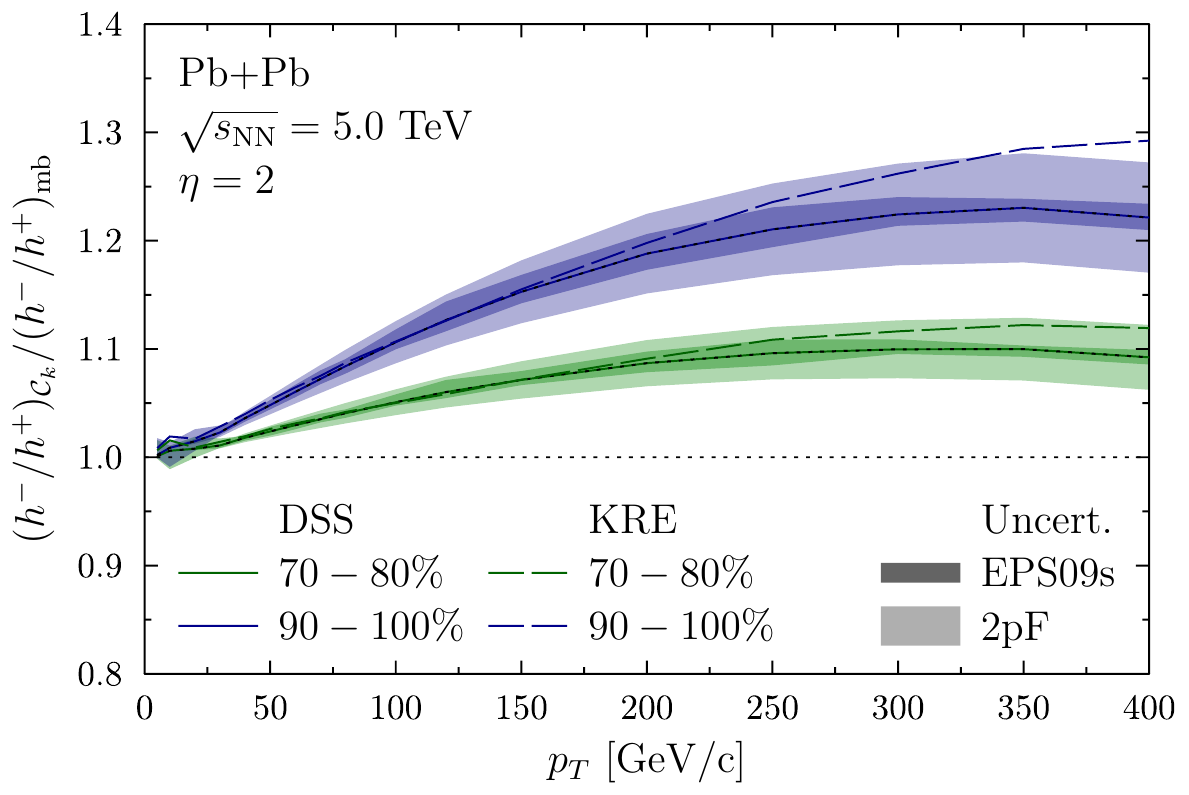}
\caption{The ratio $h^-/h^+$ for Pb+Pb collisions at $\sqrt{s_{\mathrm{NN}}} = 5.0~\mathrm{TeV}$ for $\eta = 0$ (left) and $|\eta| = 2$ (right) with \textsc{DSS} (solid) and \textsc{Kretzer} (dashed) FFs. The centrality classes 70--80~\% (green) and 90--100~\%  (blue) are normalized with the MB result. The uncertainty bands are from the 2pF distribution (light colour band) and nPDF uncertainties (dark colour band). From Ref.~\cite{HPE2016}.}
\label{fig:RnegPosHadr}
\end{figure}

The ratios $h^-/h^+$ for two centrality classes, 70--80~\% and 90--100~\%, are shown in Figure \ref{fig:RnegPosHadr} for $\eta = 0$ and $|\eta|=2$, normalized by the ratios in MB collisions. As expected, the nPDF effects and also their centrality dependence cancel out in this observable, although a small uncertainty originating from the nPDFs is still visible. This uncertainty is, however, much smaller than the NS effect and also significantly smaller than the uncertainty that originates from the 2pF distribution. The 2pF uncertainty is larger than in the case of photons since the the cross section for negative hadrons is more sensitive to the parameter variations of the neutron distribution. At $\eta = 0$ the NS effect generates an increase up to 20~\% for the $h^-/h^+$ ratio at the highest values of $p_{\rm T}$ for the most peripheral collisions. With $|\eta|=2$ a similar effect is seen already around $p_{\rm T} \approx 200~\mathrm{GeV/c}$ \cite{HPE2016}.

\section{$W$ production}

\begin{figure}[tbh]
\centering
\includegraphics[clip, trim=0pt 3pt 0pt 5pt, width=0.9\textwidth]{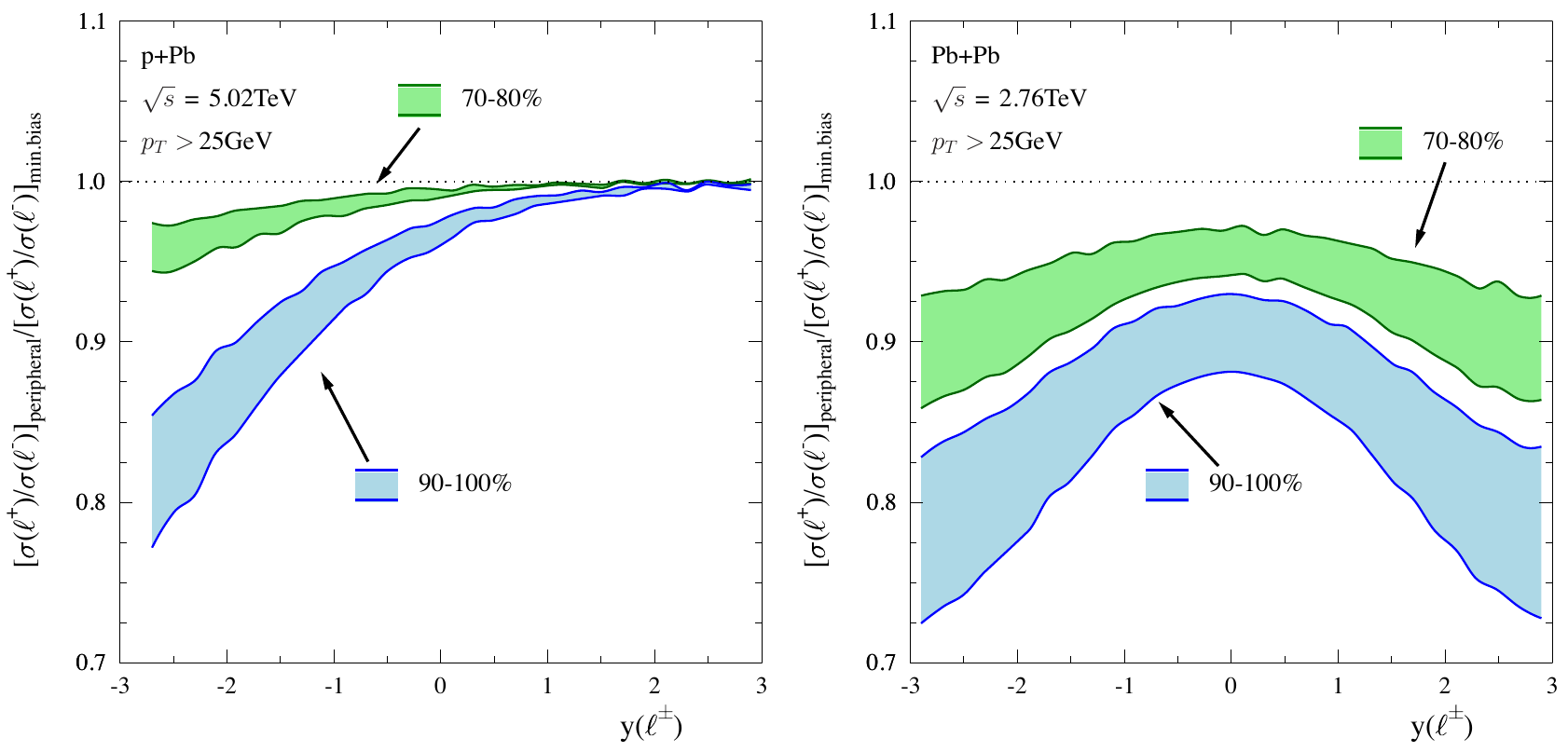}
\caption{The ratio between the positive and negative lepton cross sections from $W^{\pm}$ decays for $p_{\rm T}^{l^{\pm}}> 25~\mathrm{GeV/c}$ in 70--80~\% (green) and 90--100~\% (blue) centralities in p+Pb (left) and Pb+Pb (right) collisions, normalized with the MB result. The colour bands show the uncertainties of the 2pF distribution. From Ref.~\cite{Paukkunen:2015bwa}.}
\label{fig:W}
\end{figure}

The $W$ production is studied with charged leptons coming from $W^{\pm}\rightarrow l^{\pm}\nu$ decays. The cross sections are calculated at NLO using the \textsc{MCFM} code \cite{mcfmpage} and we consider the ratio between positive and negative lepton cross sections ($l^+/l^-$) to quantify the NS effect. The nPDF effects are not included since they are expected to cancel out (see Ref.~\cite{Paukkunen:2015bwa}), which is supported by the results for charged hadron production above. Figure~\ref{fig:W} shows the rapidity dependence of the $l^+/l^-$ ratio integrated over $p_{\rm T}^{l^{\pm}}> 25~\mathrm{GeV/c}$ for p+Pb and Pb+Pb collisions at the LHC in 70--80~\% and 90--100~\% centralities, normalized with the MB result. The NS effect is more pronounced at large rapidities where the cross sections are more sensitive to higher nuclear $x$. In general, the effects are of the similar order as in the $h^-/h^+$ ratio and also the 2pF uncertainties are of a similar size.

\section{Summary}

We have studied the impact of the NS effect to the high-$p_{\rm T}$ observables in nuclear collisions at the LHC. For the direct photons the NS effect yields only a small suppression for peripheral events, which is further masked by the nPDF uncertainties and their spatial dependence. The nPDF effects cancel out when considering the ratio between negative and positive hadrons, or the ratio between positive and negative leptons from $W$ decays, and the impact of the NS effect is more pronounced. Modifications up to 20~\% are expected, which should be large enough to be measured at the LHC if the required fine centrality binning can be performed. An intriguing possibility would be to use the NS effect to calibrate the centrality classification in different types of collisions, possibly including also the future electron-ion collisions.

\acknowledgments
I.~H. has been supported by the MCnetITN FP7 Marie Curie Initial Training Network, contract PITN-GA-2012-315877 and has received funding from the European Research Council (ERC) under the European Union's Horizon 2020 research and innovation programme (grant agreement No 668679). H.~P. was supported by the European Research Council grant HotLHC ERC-2011-StG-279579 and by Xunta de Galicia (Conselleria de Educacion) and is part of the Strategic Unit AGRUP2015/11.

\bibliographystyle{JHEP}
\bibliography{neutronSkin}

\end{document}